\documentclass[allclo]{FBSart}
\usepackage{amsfonts}
\usepackage{amssymb}

\title{On the existence of exotic and non-exotic multiquark meson states}
\author{J. Vijande$^{1,2}$\thanks{\textit{E-mail address:} 
javier.vijande@uv.es}, E. Weissman$^3$, A. Valcarce$^2$, N. Barnea$^{3,4}$}
\institute{$^1$Departamento de F\' \i sica Te\'orica e IFIC,
Universidad de Valencia, 46100 Burjassot, Spain.\\
$^2$Departamento de F\'\i sica Fundamental, Universidad de Salamanca, 37008 Salamanca, Spain.\\
$^3$The Racah Institute of Physics, The Hebrew University, 91904, Jerusalem, Israel.\\
$^4$Institute for Nuclear Theory, University of Washington, Seattle, WA 98195, USA}

\runningauthor{J. Vijande}
\runningtitle{On the existence of exotic and non-exotic multiquark meson states}
\begin{document}

\maketitle

\begin{abstract}
To obtain an exact solution of a four-body system containing two quarks and two antiquarks
interacting through two-body terms is a cumbersome task that has been tackled with 
more or less success during the last decades. We present an exact method for the study of four-quark systems based on the
hyperspherical harmonics formalism that allows us to solve it without resorting to further approximations, like 
for instance the existence of diquark components. We apply it to systems containing two heavy and 
two light quarks using different quark-quark potentials. While $QQ\bar n \bar n$ states may 
be stable in nature, the stability of $Q\bar Qn \bar n$ states would imply the existence of quark
correlations not taken into account by simple quark dynamical models.
\end{abstract}

The discoveries on several fronts~\cite{Pdg06}, of unusual 
charmonium states like $X(3872)$ and $Y(4260)$ and 
open-charm mesons with unexpected masses like
$D_{sJ}^*(2317)$ and $D^*_0(2308)$, 
have re-invigorated the study of hadronic resonances. 
Any debate on the possible multiquark structure of meson
resonances should be based on our
capability to find an exact solution of the four-body 
problem~\cite{Ade82}. Theoretical predictions
often differ because of the approximation method used.
A powerful tool to solve a few-particle system is an
expansion of the trial wave function in terms of hyperspherical harmonics (HH)
basis functions. In Ref. \cite{Vij07} a generalization of the 
HH formalism to study four-quark systems
in an exact way was presented.
Due to their actual interest and having in mind that systems with
unequal masses are more promising~\cite{Ade82}, we will center our attention on
$Q Q\bar n \bar n$ and $Q  \bar Q n\bar n$ states ($n$ stands
for a light quark and $Q$ for a heavy one). We will analyze the possible
existence of compact four-quark bound states using two standard quark-quark
interactions, a Bhaduri-like potential (BCN)~\cite{Bha81} and
a constituent quark model considering boson
exchanges (CQC) \cite{Vij05}. Both interactions
fulfill the requirement of giving a reasonable description
of meson and baryon spectroscopy. Assuming non-relativistic quantum 
mechanics we solve the four-body Schr\"odinger 
equation. The grand angular momentum
$K$ is the main quantum number in our expansion and the
calculation is truncated at some $K$ value. Further details of the numerical method can be 
found in Ref. \cite{Vij07}.

In spite of the shortcomings of the methods used to study
four-quark systems, in the past, many four-quark bound states 
have been suggested. 
To analyze their stability against dissociation,
parity and total angular momentum 
must be preserved. Additionally, $C-$parity
is a good quantum number for $c\bar c n\bar n$
and the Pauli principle must be fulfilled in the $cc\bar n\bar n$ case.
The corresponding thresholds can be computed by adding the meson masses
of the dissociation channel. 
Four-quark states will be stable under strong 
interaction, and therefore very narrow, if their total 
energy lies below all allowed two-meson thresholds. 
Sometimes, results of four-quark calculations have been directly compared to
experimental thresholds. In this case 
one could misidentify scattering wave functions as bound states.
When they are referred to the thresholds within the same model, theoretical 
predictions do not imply an abundance of multiquark states in the data.

\begin{table}[hbt]
\beforetab
\begin{tabular}{cp{4mm}cccp{4mm}ccc}
\firsthline
&&\multicolumn{3}{c}{CQC}&&\multicolumn{3}{c}{BCN}\\
\midhline
$K$     && E   & $P_{11}$&$P_{88}$&& E    & $P_{11}$& $P_{88}$    \\
\midhline
18      && 3791&  0.9962 & 0.0038&& 3840  &0.9995 & 0.0005 \\
20      && 3786&  0.9968 & 0.0032&& 3822  &0.9996 & 0.0004 \\
22      && $-$ &  $-$    &  $-$  && 3808  &0.9997 & 0.0003 \\
\midhline
$J/\psi\,\omega\vert_S$ && 3745  & 1 & 0 && 3874 & 1 & 0  \\
$\chi_{cJ}\,\eta \vert_P$ && 4281  & 1 & 0 && 3655 & 1 & 0  \\
\lasthline
\end{tabular}
\aftertab
\captionaftertab[]{Energy (MeV) and probability of the different 
color components for the $c\bar c n \bar n$ $J^{PC}=1^{++}$ 
both for CQC and BCN models. 
The last rows indicate the lowest theoretical two-meson thresholds.
$P_{11}$ ($P_{88}$) stands for
the probability of singlet-singlet (octet-octet) color 
components.}
\label{t2}
\end{table}

Once the method has been established, we concentrate on the $c\bar c n\bar n$ systems 
as a potential structure 
for the $X(3872)$. To make the physics clear we compare 
with the $cc \bar n \bar n$ system. In particular, we focus 
on the $J^{PC}=1^{++}$ $c\bar c n\bar n$ and 
$J^{P}=1^{+}$ $cc\bar n \bar n$ quantum numbers to illustrate
their similitudes and differences. 
A complete study of all the quantum numbers 
have been reported in Ref. \cite{Vij07}. 
For the $c\bar c n\bar n$ system, independently
of the quark-quark interaction, the system evolves to a 
well separated two-meson state, see Table \ref{t2}. This is clearly seen 
in the energy, approaching the corresponding two free-meson threshold, 
but also in the probabilities of the 
different color components of the wave function.
Comparing the theoretical predictions with the experimental threshold, 
$M_{J/\psi\,\omega\vert_S} = 3879.57\pm0.13$ MeV,  
one could be tempted to claim for the existence of a bound state. However, the experimental
threshold is not reproduced by the effective Hamiltonians. Thus, in any manner one can claim for the existence
of a bound state.  Similar conclusions are drawn for 
all quantum numbers of this system. A completely different behavior is observed in the case 
of $J^P=1^+$ $cc\bar n\bar n$.
The energy quickly stabilizes below the lowest theoretical thresholds (3937 MeV for CQC and 3906 for BCN), 
being the results obtained for $K_{max}=24$ completely converged, 3861 MeV for CQC and 3900 for BCN. Besides, 
the radius is also stable and is smaller than the
sum of the radius of the two-meson threshold. We obtain $r_{4q}=0.37$ fm
compared to $r_{M_1}+r_{M_2}= 0.44$ fm. 

It is thus important to realize that a bound
state should be pursued not only by looking at the energy, but also
with a careful analysis of the radius and probabilities. This detailed 
analysis allows us to distinguish between 
compact states and meson-meson molecules \cite{Jaf05}
and it does consider the contribution of all meson-meson channels
to a particular set $J^{PC}$ of quantum numbers.
Inherent to our discussion is a much richer decay spectrum
of compact states due to the presence of octet-octet color
components in their wave function.

Let us notice that
there is an important difference between the two physical systems studied.
While for the $c\bar c n\bar n$ there are two allowed physical {\it decay channels},
$(c\bar c)(n\bar n)$ and $(c\bar n)(\bar c n)$, for the $cc\bar n\bar n$ 
only one physical system contains the possible final states, $(c \bar n)(c\bar n)$. 
This has important consequences if both systems (two- and four-quark
states) are to be described within the same two-body Hamiltonian,
the $c \bar c n \bar n$ will hardly present bound states, because the system
will reorder itself to become the lightest two-meson state, either
$(c\bar c)(n\bar n)$ or $(c\bar n)(\bar c n)$. In other words,
if the attraction is provided by the interaction between
particles $i$ and $j$, it does also exist in the asymptotic
two meson state reflecting this attraction. This may not happen
for the $c c\bar n\bar n$ if the interaction between, for example,
the two quarks is strongly attractive. In this case there is no asymptotic two-meson
state with such attraction, and therefore the system will bind. 

Therefore, our conclusions can be made more general. If we have an $N$-quark
system described by two-body interactions in such a way that there exists
a subset of quarks that cannot make up a physical subsystem, then one may expect
the existence of $N$-quark bound states by means of central two-body potentials.
If this is not true one will hardly find $N-$quark bound states \cite{Lip75}.

\begin{acknowledge}
This work has been partially funded by Ministerio de Ciencia y 
Tecnolog\'{\i}a under Contract No. FPA2007-65748, and by 
Junta de Castilla y Le\'{o}n under Contract No. SA016A07. 
\end{acknowledge}


\begin{thebibliography}{9}

\bibitem{Pdg06} W.-M. Yao {\it et al.},
                        J. Phys. G {\bf 33} 1 (2006).

\bibitem{Ade82} J.P. Ader, J.-M. Richard, and P. Taxil,
                Phys. Rev. D {\bf 25}, 2370 (1982);
                L. Heller and J.A. Tjon,
                Phys. Rev. D {\bf 32}, 755 (1985); {\it ibid} {\bf 35}, 969 (1987).

\bibitem{Vij07} N. Barnea, J. Vijande, and A. Valcarce,
                Phys. Rev. D {\bf 73}, 054004 (2006);
		J. Vijande, E. Weissman, A. Valcarce, and N. Barnea,
                Phys. Rev. D {\bf 76}, 094022 (2007).

\bibitem{Bha81} R.K. Bhaduri, L.E. Cohler, and Y. Nogami,
                Nuovo Cimento {\bf A65}, 376 (1981).

\bibitem{Vij05} J. Vijande, F. Fern\'andez, and A. Valcarce, 
                J. Phys. G {\bf 31}, 481 (2005).

\bibitem{Jaf05} R.L. Jaffe, 
                        Phys. Rep. {\bf 409}, 1 (2005);
                        hep-ph/0701038.

\bibitem{Lip75} H.J. Lipkin,
                        Phy. Lett. {\bf 58B}, 97 (1975).

\end{thebibliography}
\end{document}